\documentclass[12pt, letterpaper]{article}

\usepackage{amsmath,amssymb}
\usepackage{cite}
\usepackage{fancyhdr}
\usepackage[top=1in, bottom=1in, left=1in, right=1in]{geometry}
\usepackage{graphicx}
\usepackage{hyperref}

\numberwithin{equation}{section}
\date{}

\begin{document}
\title{{\rm\footnotesize \qquad \qquad \qquad \qquad \qquad \ \qquad \qquad \qquad \ \ \ \ \ \                      RUNHETC-2024-31
}\vskip.5in    Physics, Philosophy, "Observers" and Multiverses}
\author{Tom Banks\\
NHETC and Department of Physics \\
Rutgers University, Piscataway, NJ 08854-8019\\
E-mail: \href{mailto:tibanks@ucsc.edu}{tibanks@ucsc.edu}
\\
\\
}

\maketitle
\thispagestyle{fancy} 

\begin{abstract} \normalsize We comment on the fact that certain mathematical models that have been proposed in the quantum gravity literature, may not be subject to experimental checks, even if they turn out to be mathematically well defined.  This means that they would be indistinguishable from other models, with a smaller number of quantum states.  These considerations follow from very general properties of quantum measurement theory and classical gravity.  \noindent  \end{abstract}


\newpage
\tableofcontents
\vspace{1cm}

\vfill\eject
\section{Introduction}

At the moment, science has no complete understanding of the emergence of life from non-living matter, let alone of the emergence of what we call conscious self awareness from more primitive forms of life.  In particular, we do not really know the role that language plays in what we call consciousness or intelligence\footnote{See for example, octopi.}.  What we do know is that language played an extremely important role in the development of conscious awareness and intelligence for homo sapiens on the planet Earth.  We also know that language is the most important tool our species uses to further our genetically imprinted urges toward lying, cheating and fantasizing.  This was perhaps said best by Francis Bacon at the dawn of the re-invention of science in western Europe. 

The trouble, Bacon explained, is that everything, even the framing of experiments, begins with language, with words; and words have a fatal tendency to substitute themselves for the facts they are supposed merely to report or reflect. While men believe that their reason governs words, in fact words react on the understanding; that is, they shape rather than serve rationality. Even precise definitions, Bacon lamented, don't help because the definitions themselves consist of words, and those words beget others and as the sequence of hypotheses and calculations extends itself, the investigator is carried not closer to but ever further way from the independent object he had set out to apprehend.

In Bacon's mind the danger of words going off on their own unconstrained-by-the-world way was but one example of the deficiencies we have inherited from the sin of Adam and Eve. In men's love of their own words (and therefore of themselves), he saw the effects "of that venom which the serpent infused and which makes the mind of man to swell." As an antidote he proposed his famous method of induction, which mandates very slow, small, experimental steps; no proposition is to be accepted until it has survived the test of negative examples brought in to invalidate it.

In the evolution of scientific thought since Bacon's time we have come to accept two different methods of scientific reasoning as acceptable definitions of truth.  The first is essentially the Baconian one: everything must be subjected to experimental test.  This has been modified by the notion of what we now call effective theories.  We've found, over and over again, that the fundamental conceptual foundations of our theories of physics are wrong, despite their agreement with experiment over a wide range of conditions.  The reason for this is that the more correct theory, which replaces them in regimes where they fail, is based on concepts unfamiliar to our everyday intuition.  It is more complex, and is well approximated by the "wrong" effective theory, within a certain range of conditions.  So we never throw out the wrong theory, just accept its limitations.

The second accepted definition of truth is mathematical rigor.  Once a proof has been properly vetted, all mathematicians will be able to follow it and verify its validity by following a set of well defined rules.  There also seems to be a mysterious\cite{wigner} way in which mathematical concepts "inhabit" the physical world, so that we can use the tools of mathematics to prove truths about "reality".  However, here we are entering into treacherous territory and are in danger of crossing the boundary between science and philosophy that Francis Bacon tried so hard to establish.   Mathematics quite obviously has many things in it that have nothing to do with the real world.  We don't yet know what they all are, and it would be a mistake to jump to conclusions.  If some genius in the 19th century had realized that complex linear algebra defined a new kind of probability theory, in which Bayes' rule of conditional probabilities was not true, that person would have been laughed out of the university and sent to an insane asylum.  If you had asked me in graduate school whether I thought there would ever be any use in physics for a concept called "modular tensor categories" (which didn't yet exist) I would have said no.  The mathematicians from whom I learned category theory called it "abstract nonsense", a technical term.  Nonetheless, it's clear that "there are more things in mathematicians' philosophy than are dreamed of in heaven and earth". 

In this brief note I want to note two popular subjects in contemporary high energy theoretical physics, where I think mathematical philosophizing has ignored the Baconian injunction to submit hypotheses to careful tests.  I've discussed one of these at length before\cite{landscape} but recently attended a conference where it was clear that not all of those points had come across.  The other is the recently fashionable discussion of "observers" of finite regions of space-time in terms of the crossed product construction of Type II factors from Type III factors in algebraic quantum field theory.  I will begin with the latter.

I much prefer the terminology "detector" to "observer", since it avoids all of the confusion about the connection of quantum mechanics to consciousness that has generated so much nonsense in popular literature\footnote{See the novel {\it Dark Matter} by Blake Crouch, and the even sillier TV adaptation on Apple+ , for a recent example.} .  As first appreciated by von Neumann, and made more precise by work on decoherence in the 1960s and 1970s, a detector is a large quantum system, with many collective variables called {\it pointers}.  Each pointer corresponds to a large subsystem with of order $e^N$ states and $N$ is a reasonably large power of $10$.  One can think of $N$ as the number of atoms in the pointer.  For concreteness we imagine the system the detector is supposed to probe is made up of q-bits and that each pointer variable is two valued.  
The commutators of the pointer variables are proportional to inverse powers of $N$, and the interference terms between different quantum histories of the pointer variables are exponentially small in $N$.  If this is the case, the quantum predictions for probabilities of observations made on the pointer variables obey stochastic differential equations, with small fluctuation terms and the probabilities satisfy Bayes' conditional probability rule with accuracy exponential in $N$..  If there is one independent pointer variable per q-bit in the system, then the detector can make a complete measurement of the quantum state of the system.  A measurement is a unitary operation that creates a completely entangled state between the q-bits of the system and the two valued pointer variables of the detector.  One then uses Bayes' rule to make conclusions about interactions between the pointer variables of the detector and those of other macroscopic systems.  The rules of classical probability theory apply and apart from the usual worries about not being able to do an infinite number of re-runs of each experiment we have a completely comprehensible theory of prediction of the quantum behavior of the system being measured.  

It's immediately obvious that a detector cannot detect everything about the quantum state of a universe, or part of the universe that contains itself, because the pointer variables are by definition insufficient to differentiate the quantum states of the detector. However, simple things that we know about how to construct mathematical models of detectors, and about the classical physics of gravitation, tell us that the {\it a priori} constraints on the notion of a detector in a theory of quantum gravity are much more severe than the above self evident statements.  

Mathematical models of systems with pointer variables, which apply to the real world, are all cut off versions of quantum field theories, most rigorously constructed on a lattice.  The pointers are made from averages of local fields (typically conserved densities) over large regions.  However, it has been known for a long time\footnote{I first heard this argument from G. 't Hooft some time in the 1980s, but it surely pre-dates that.} that most of the states that a quantum field theory assigns to a large space-time region have a large semi-classical backreaction. If we use the expectation value of the stress tensor in those states to compute a gravitational field, it forms a black hole larger than the original region.  Imposing some kind of cutoff to avoid such black holes in a fixed size causal diamond, one finds that the logarithm of the number of remaining states scales like
\begin{equation} (\frac{A_{\diamond}}{G_N})^{\frac{d-1}{d}} , \end{equation} where $A_{\diamond}$ is the volume of the largest spacelike $d-2$ surface on the boundary of the diamond.  In an important paper\cite{CKN} Cohen, Kaplan and Nelson showed that omitting all of the offending states from QFT did not affect the agreement between QFT and the most precise experiments ever done.  Further refinements of their result can be found in\cite{draperetal} and indicate that it is unlikely any disagreement will ever be found, because the effect can be pushed away by different choices of the field theory cut off. 

On the other hand, a variety of arguments\cite{coventropy} suggest that the actual entropy in a diamond is given by $\frac{A_{\diamond}}{4G_N}$.  This means that, in principle, the number of quantum states in any detector is smaller by a factor of order 
\begin{equation} e^{- (\frac{A_{\diamond}}{4G_N})^{\frac{1}{d}}} , \end{equation} than the number of states in the diamond, while the number of pointer variables is exponentially smaller than the number of states in the detector.  

The situation in de Sitter space is demonstrably worse, and if the conjectures of\cite{fs}\cite{bousso}\cite{carlip}\cite{solo}\cite{BZ} are correct, any causal diamond for non-negative cosmological constant, and any diamond smaller than the AdS radius for negative c.c. behaves quite analogously.  For dS space of radius $R$, localized objects with mass $M$ correspond to constrained states where of order $MR$ (in units where $\hbar = 1$) q-bits are frozen relative to the empty diamond density matrix.  Furthermore, most of the states of mass $M$ are black holes, which do not have a lot of pointer variables and cannot serve as useful detectors.  So, at least in dS space, the more complex and sophisticated a detector we build, the smaller the subspace of the full Hilbert space we are allowed to probe.  This problem comes on top of the restrictions we've already discussed. 
In light of these {\it a priori} constraints coming from semi-classical restrictions on the validity of the QFT description of the space of states in a localized region of space-time, it appears that any attempt to model a detector by a quantum mechanical degree of freedom with a continuous spectrum grossly overcounts the actual capabilities of any physically realizable detector.  

The mathematical basis for that model of detectors was the crossed product construction of Type $II_{\infty}$ algebras from the Type $III_1$ algebras assigned to causal diamonds by quantum field theory.  It should be noted that this construction, although it allows certain operators to have finite traces, actually enlarges the Type $III_1$ algebra, which certainly violates the spirit of what the arguments about black holes lead us to expect of the modification of QFT by gravity.  It has been argued that the Type $II_1$ truncation of the crossed product that has been proposed for dS space\cite{CLPW} only contains (in perturbation theory) the non-local BRST invariant sub-algebra of field theory operators.  If this is true, it is at least consistent with the intuition that gravitational effects drastically reduce the number of physical operators.

The question at issue here is whether any of these objects, which apart from the non-renormalizability of gravity, are mathematically well defined in perturbation theory, actually correspond to things that a physical detector in real dS space could detect.  The longest lived physical detectors in a universe like our own travel along some time-like trajectory in a Schwarzschild dS space-time.  The Schwarzschild-dS space is associated with some local group of galaxies, the trajectory of whose center of mass is decohered by the many quantum states of the galaxies' constituents.  This trajectory is not exactly a geodesic because of the interaction of the local group with Gibbons-Hawking radiation, even after all other macroscopic bodies have gotten microscopically close to the local group's cosmological horizon. If one considers the center of mass of the local group to be the position of a relativistic particle, it would be spread around the horizon of any geodesic's causal diamond in a {\it relatively} short time by its own quantum fluctuations or the effect of interaction of Gibbons-Hawking radiation and redshifting of quasi-normal modes.  This is probably irrelevant.  The center of mass is entangled with all of the constituents of the local group, so it will define its own cosmological horizon until it collapses into a black hole (the latter time is the bound on the utility of our detector, which is some macroscopic subsystem in the local group) and radiates its degrees of freedom back to thermal equilibrium.  

As we've argued above, our detector can only access a finite amount of quantum information, smaller by multiple exponential factors than the Gibbons Hawking entropy of dS space.  There's certainly no meaning to talking about an infinite dimensional algebra of quantum operators that are accessible to measurement, even in principle.
The Type $II_1$ proposal for dS space or other finite area causal diamonds is thus a part of philosophy, and not physics.  Even finite dimensional models which talk about a Hilbert space whose size is determined by the Gibbons Hawking entropy formula will only be amenable to partial experimental verification.

Although this is tangential to the point of the present paper, I want to repeat a point that was made in\cite{tbpddS}.  Causal diamonds in QFT do not have density matrices because the singularity of field theory commutators on the light cone implies an infinite entanglement entropy.  It also implies an infinite uncertainty $(\Delta K)^2$ for the modular Hamiltonian.  With {\it any} regularization scheme the ratio of these two quantities is finite and non-zero.  There's a class of regularization schemes that puts a UV cutoff on the momenta tranverse to the light cone, which obtains $(\Delta K)^2 = \langle K \rangle $, essentially because they preserve a near horizon Virasoro symmetry\cite{carlip}\cite{solo}.  A beautiful regularization scheme\cite{CHM} that preserves higher dimensional conformal invariance, relates this ratio to a coefficient in the two point function of the stress tensor\cite{perl}.  Remarkably, for all CFTs that have large radius holographic duals, the coefficient is again equal to $1$\cite{deboeretal} in agreement with a replica calculation\cite{VZ2}.   This is the value predicted quite generally in quantum gravity by the arguments of\cite{carlip}\cite{solo} as generalized by\cite{BZ}. The result $(\Delta K)^2 = \langle K \rangle $ has also been obtained recently for dS space by a replica trick\cite{tbpddS}. For large black holes in $AdS_d$, the coefficient is instead $d-2$ because of the tensor network structure of the quantum theory and the dominance of the entropy by compressible modes of the tensor network.  In all cases the expectation value and fluctuation of the modular Hamiltonian are rendered finite simultaneously and are proportional.  {\it The crossed product construction and its Type $II_1$ deformation fail to give results resembling these.}  In the crossed product one makes the entropy finite but the fluctuation is infinite.  The hypothesis that the density matrix is the unit matrix (originally made in\cite{tbds}\cite{wfds}\cite{bousso}) gives zero fluctuation.  

\section{Multiverses}

String theory and ideas about Eternal Inflation\cite{susskind}\cite{BP}\cite{linde}\cite{guth} have popularized the idea of a {\it multiverse}: the possibility that the theory of quantum gravity actually predicts a plethora of very different macroscopic space-times and allows for quantum mechanical tunneling transitions between them.  I have written extensively about arguments showing that there is no positive evidence for these claims and lots of negative evidence against them\cite{tblandscape}.  There is also a peculiar and completely unfounded conflation of these ideas with the Many Worlds interpretation of quantum mechanics.  That purely philosophical point of view is a logically distinct set of ideas, which has no mathematical or physical connection to string theory or eternal inflation.   In this note I will not regurgitate or even summarize the arguments against the mathematical existence of the landscape of string theory or some of the mistaken interpretations of the Coleman Deluccia tunneling paper.  Instead we ask the question of whether a theory of a multiverse (of the type actually written about in scientific papers as opposed to science fiction movies) belongs to physics or philosophy.

Perhaps the primary reason a real physicist would be interested in a multiverse would be to explain the values of disturbingly small or large constants in their theory of nature. A. Albrecht once said that "The person who has the smallest number of {\it anthropically} determined constants in their theory wins."  At present we have (a number of) perfectly plausible explanations for all anomalously small numbers in physics apart from the cosmological constant. The suggestion that the c.c. be determined anthropically was originally made by\cite{Barrow}.  Field theory models that tried to implement this were invented in\cite{davies}\cite{tbcc}\cite{lindecc} and Weinberg\cite{weinbergcc} argued that the anthropic bound was only two orders of magnitude higher than the observational bound at the time his paper was published\footnote{In fact, when I presented the model of\cite{tbcc} at a seminar in Austin in 1985, Weinberg argued to me, and to the astrophysics group at the University of Texas, that the anthropic argument missed by many orders of magnitude.  The argument was simple: as long as the c.c. is smaller than the energy density at the time galaxies begin to form, it doesn't interfere with the formation of at least a single galaxy, which is all that one needs for life.  At that time, Weinberg was missing the factor of $(\frac{\delta \rho}{\rho})^3 \sim 10^{-15}$, which goes into the calculation of the actual bound. At the time, I knew nothing about galaxy formation, and took the word of an expert.}.   The apparent necessity of explaining the value of the c.c. by anthropic arguments has driven much of the work on the string landscape, in particular the most successful idea\cite{BP}.  

While I have no real objection to the basic idea of solving the c.c. problem by anthropic arguments, and even have a version of such an argument that fits with my own understanding of what the c.c. is\cite{tbwfcc}, I've come to believe that we should think of such arguments as part of philosophy rather than physics.  The point is that any such argument postulates some version of a multiverse, a large system that can have different states or different subsystems, each of which behaves like a universe.  Some of them will be similar to our own but with different values of the parameters that go into our fundamental description of nature.  Then one wants the theory to determine some kind of probability distribution for different values of the constants, on top of which we want to put constraints having to do with the existence of intelligent life.

Here we immediately run into the problem alluded to in our first sentence.  We do not even know how to explain life, let alone consciousness and intelligence, based on the physics that we know.  How can we ever hope to do it for all of the mathematical possibilities that field theory and string theory have revealed to us?  If we can't, what right do we have to use anthropic reasoning based on our very narrow understanding of what life is?   Weinberg's answer to that would be that galaxy formation is a purely gravitational problem and could be argued to be necessary to almost any conceivable form of life.  I find that answer reasonable.  However, beyond that, the utility of any particular model of a multiverse seems questionable from a scientific point of view.  A lot of the literature on the subject involves itself with processes that happen in various hypothetical mathematical models on time scales that can easily be much longer than the time for our local group of galaxies to collapse into a black hole.  They're therefore unlikely to have anything to do with any observation made by any conceivable detector in the local group, or any other local group in the universe.  Clearly these discussions are a part of philosophy, rather than physics.  Abstract discussions of whether we are typical of all possible intelligent observers in all possible states of a multiverse whose details we will never be able to probe experimentally, would have limited scientific interest even if they followed rigorously from a uniquely defined axiomatic theory of quantum gravity.  Given the current state of our knowledge, we seem to have many consistent independent mathematical models of quantum gravity. In most of those we cannot prove the impossibility of intelligent life existing. It would seem that the best we can hope for is to say that there is some sort of multiverse model, which explains the value of the c.c., and we are unlikely to find it.  

\section{Conclusions}

Human beings, like our mechanical progeny generative AI, are prone to fantasizing about things that do not actually exist in the physical world.  Even the rigorous rules of mathematics allow us to build constructs that do not correspond to reality.  The inventors of the scientific method insisted that the only way to keep this tendency in check was to submit our theoretical models to repeated experimental checks.  The combined constraints of quantum measurement theory and semi-classical black hole physics appear to put {\it a priori} limitations on the extent to which theoretical models of quantum gravity can be subjected to such checks.  In this note we've explored two examples of such constraints.  No experiment in a finite area causal diamond can come close to completely checking the predictions of even a finite dimensional quantum model of the diamond's operator algebra.   And while the puzzling value of the cosmological constant cries out for an explanation in terms of a multiverse and "environmental selection", no detailed model of a multiverse will ever be checked by experiment.
As time goes on, we will undoubtedly encounter other issues where our imagination as theoretical physicists wants to take us places where the laws of physics prevent us from ever going.

\begin{center}

{\bf Acknowledgments }

 The work of T.B. was supported by the Department of Energy under grant DE-SC0010008. Rutgers Project 833012.  
 \end{center}




\end{document}